\documentclass[a4paper,twoside,10pt]{article}
\usepackage{graphicx,publaob}

\def\papertype{Contributed paper}
\newcommand{\OIII}{[O\,{\sc iii}]}

\newcommand{\SII}{[S\,{\sc ii}]}
\newcommand{\Halpha}{H${\alpha}$}
\def\p0{\phantom{0}}
\def\papertype{Original Scientific Paper}

\def\arcmin{\hbox{$^\prime$}}

\begin{document}

\title{SUPERNOVA REMNANTS IN THE MAGELLANIC CLOUDS}

\authors{Miroslav D. Filipovi\'c and Luke M. Bozzetto}

\address{Western Sydney University, Locked Bag 1797, Penrith South DC, NSW 1797, Australia}
\Email{m.filipovic}{westernsydney.edu}{au}

\markboth{SUPERNOVA REMNANTS IN THE MAGELLANIC CLOUDS}{M. D. FILIPOVI\'C and L. M. BOZZETTO}

\abstract{We present initial results of an ongoing study of the supernova remnants (SNRs) and candidates in the Magellanic Clouds. Some 108 objects in both Clouds are considered to be either an SNR or a reliable candidate. This represents the most complete sample of known SNRs in any galaxy. therefore, this study allows us to study SNR population properties such as the size and spectral index distribution. Here, we also show 12 known Large Magellanic Cloud SNRs from type~Ia SN explosions and briefly comment on their importance.
}

\section{INTRODUCTION}

The Magellanic Clouds (MCs), with their low foreground absorption and relative close proximity of 50~kpc and 60~kpc (di Benedetto, 2008), offers the ideal laboratory for the study of a complete sample of supernova remnants (SNRs) in great detail. The proximity enables detailed spatial studies of the remnants in the sample, and the accurately known distance allows for analysis of the energetics of each remnant. In addition, the wealth of wide-field multi-wavelength data available, from radio maps to optical emission-line images and broad band photometry to global X-ray mosaics, provides information about the contexts and environments in which these remnants are born and evolve.

A complete sample of SNRs provides the ability to study the global properties of SNRs, in addition to carrying out detailed analysis on the subclassed (e.g., sorted by X-ray and radio morphology or by progenitor SN type). Toward this goal, we have been identifying new SNRs using combined optical, radio and X-ray observations.

A distinguishing characteristic of SNRs in radio frequencies is their well-established predominantly non-thermal continuum emission. Collectively, SNRs have a radio spectral index of $\alpha\sim-0.5$ (defined by $S\propto\nu^\alpha$), although may vary quite a lot due to the wide variety of SNRs, differing environments, and stages of evolution (Filipovi\'c et al. 1998). On one side, younger and very old remnants can have a spectral index of $\alpha\sim-0.8$, while mid-to-late-age remnants, or those which harbour a Pulsar Wind Nebulae (PWN), tend to have flatter radio spectra with $\alpha \sim -0.2$. As one of the most energetic class of sources in the Universe, these objects greatly impact the structure, physical properties and evolution of the interstellar medium (ISM). Conversely, the interstellar environments in which SNRs reside, will heavily affect the remnants' evolution. Here, we report on radio-continuum observations of the most up-to-date sample of MCs SNRs and SNR candidates, consisting of 83 objects in the Large Magellanic Cloud (LMC) and 25 in the SMC.

\section{MAGELLANIC CLOUDS SURVEYS}

There are several present-generation multi-wavelength surveys that we used in our study. Radio-continuum surveys were predominately based on observations from the Australian Telescope Compact Array (ATCA), including the 20-cm mosaic by Filipovi\'c et al. (2002); Payne et al. (2004); Hughes et al. (2007); Wong et al. (2011a,b) as well as the 6-cm and 3-cm mosaics published by Dickel et al. (2005); Crawford et al. (2011); Wong et al. (2012). In addition, a 36~cm Molonglo Synthesis Telescope (MOST) mosaic image (as described in Mills et al. 1984) was used. We note that the former surveys (i.e., those from the ATCA) included a zero-spacing measurement from the single 64-m Parkes dish (Filipovi\'c et al. 1995, 1997), while the latter MOST image did not. This resulted in missing short spacings at $\lambda$=~36 cm, and therefore, the potential for missing flux may be an issue, especially for larger remnants.

Over the past several years we performed a number of X-ray surveys of both MCs using the \textit{ROSAT}, \textit{Chandra} and \textit{XMM-Newton} observatory (Haberl et al.~2012; Maggi et al.~2016). For example, the \textit{XMM-Newton} LMC large project comprises 25~ks observations of 70 fields, which together with archival data cover an area of about 10 square degrees. The \textit{XMM-Newton} surveys provide a unique data set to investigate the X-ray source populations of the MCs including SNRs and candidates. Our latest search for the new LMC SNRs resulted in a discovery of 4 such objects (Maggi et al. 2014). The comprehensive review on the SMC X-ray sample can be found in Filipovi\'c et al. (2008) and Owen et al. (2011). Additional X-ray data was sourced from the Chandra Supernova Remnant Catalog.

Data from the Magellanic Cloud Emission Line Survey (MCELS) was used throughout this work. This survey was carried out at the 0.6~m University of Michigan/CTIO Curtis Schmidt telescope (Smith et al. 2006). Both MCs were mapped in narrow bands corresponding to \Halpha, \OIII\ ($\lambda$=5007\,\AA), and \SII\ ($\lambda$ = 6716,\,6731\,\AA), plus matched red and green continuum bands. Our own spectroscopic surveys of the LMC (Payne et al. 2008) and SMC (Filipovi\'c et al. 2005; Payne et al. 2007) SNR sample are mainly taken with the SAAO 1.9-m and MSSSO 2.3-m telescope. 

We use MCs infrared data from the \textit{Spitzer} Space Telescope and {\it Herschel} surveys (Laki\'cevi\'c et al. 2015). The Multiband Imaging Photometer were used for \textit{Spitzer} (MIPS) (24, 70 and 160~$\mu$m) and with SPIRE (Spectral and Photometric Imaging Receiver) at 250, 350 and 500~$\mu$m and with Photodetector Array Camera and Spectrometer at 100 and 160~$\mu$m.

\section{RESULTS}

We present size/diameter and radio spectral index distribution studies of SNRs from both Clouds as well as an initial study of 12 type~Ia SNRs from the LMC. We point to a number of MCs SNRs studies that our group did over the past few years. Namely, Boji{\v c}i{\'c} et al.\ (2007), {\v C}ajko et al.\ (2009), Crawford et al.\ (2010), Bozzetto et al.\ (2010), Owen et al.\ (2011), Grondin et al.\ (2012), Bozzetto et al.\ (2012a,b,c), De Horta et al.\ (2012), Kavanagh et al.\ (2013), Bozzetto et al.\ (2013), Brantseg et al.\ (2014), Bozzetto et al.\ (2014a,b,c), De Horta et al.\ (2014), Warth et al.\ (2014), Crawford et al.\ (2014), Reid et al.\ (2015), Kavanagh et al.\ (2015a,b,c), Maitra et al.\ (2015), Roper et al.\ (2015), Bozzetto \& Filipovi{\'c} (2015) and Kavanagh et al.\ (2016). 

 \subsection{Size of Magellanic Clouds SNRs}

\begin{figure}[h!]
\vspace{-3mm}
  \begin{center}
     \includegraphics[trim=0 0 0 0, width=.485\textwidth]{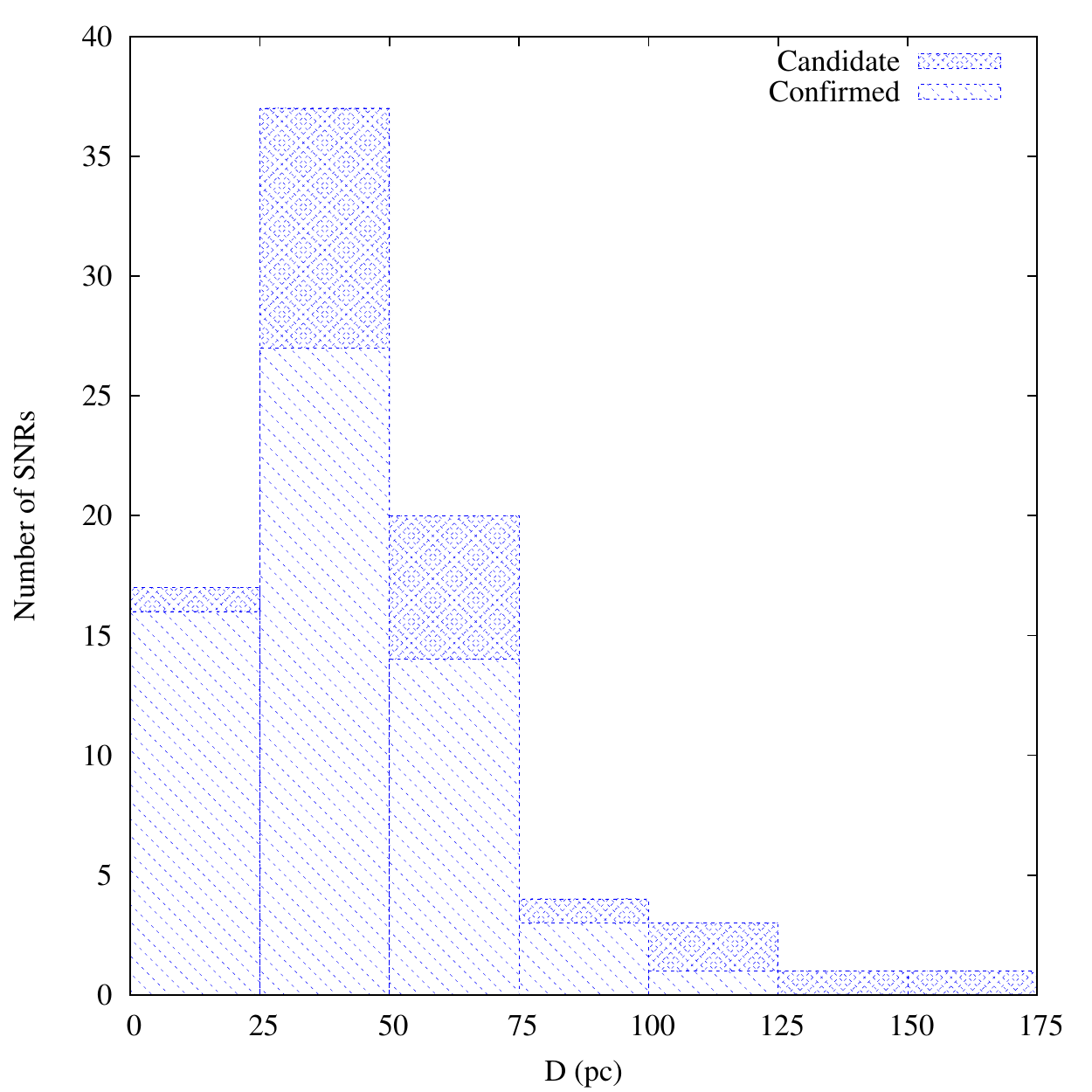}
     \includegraphics[trim=0 0 0 0, width=.485\textwidth]{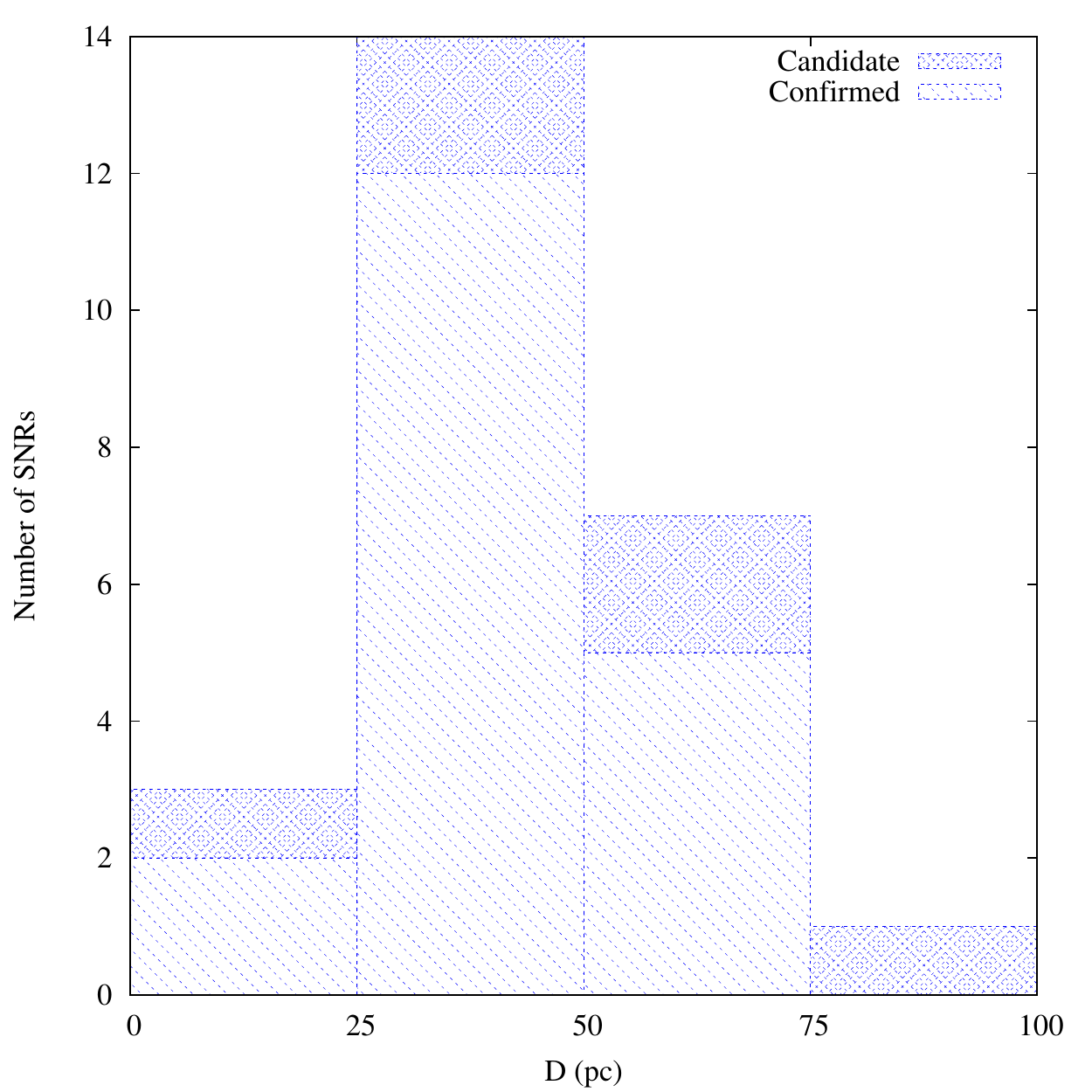}
  \end{center}
  \vspace{-8mm}
  \caption{The size distribution for SNRs in the LMC (left) and SMC (right).}
  \label{fig:sd}
\end{figure}

To measure the extent of the LMC SNR population, an ellipse was fitted to delineate the bounds of emission from all confirmed and candidate SNRs in this study. A multi-wavelength approach was used such that the given size takes into account the optical, radio and X-ray emission. The resulting histograms showcasing this data is displayed in Fig.~\ref{fig:sd}, where the diameter is the average of the major and minor axes. 

The mean diameter of {LMC} {SNRs} was is estimated to be {39.4~pc}, with a standard deviation (SD) of {22.2~pc} for some 61 confirmed {SNRs}. The entire LMC sample of 83 SNRs has somewhat larger size of {44.8~pc} with a SD of {28.1~pc}. These values are moderately larger than the value found for M\,83 in a study by {Dopita et al. (2010)}, which showed a mean diameter of 22.7~pc (SD=10.3~pc) for the 47 measured remnants. Also, in a previous study of the {SMC}, {Filipovi{\'c} et al. (2005)} found a mean diameter of the SMC sample to be $\sim$30~pc. They noted that such a value indicate that most of the remnants are in the adiabatic/Sedov evolutionary stage. The results of this study are more inline with the those found by {Long et al. (2010)} in their study of M\,33, finding a median of 44~pc and by {Lee et al. (2014)} for M\,31, which showed a strong peak at $D=48$~pc. 

The SMC SNR sample diameter was measured in similar way. Taking a 1-D slice of the remnants major and minor axes and measuring the extent at the 3$\sigma$ noise level. The mean value of the SMC was estimated to be 42.6~pc with a SD of 15.8~pc for the confirmed 19 remnants, and 44.5~pc with a SD of 17.8~pc for the sample inclusive of candidate remnants (25). With the larger sample size in comparison to the earlier study by {Filipovi{\'c} et al. (2005)}, we find both of these diameter estimates are significantly larger than the previous measurement of 30~pc.

 \subsection{Spectral Index of Magellanic Clouds SNRs}

\begin{figure}[h!]
\vspace{-3mm}
  \begin{center}
     \includegraphics[trim=0 0 0 0, width=.485\textwidth]{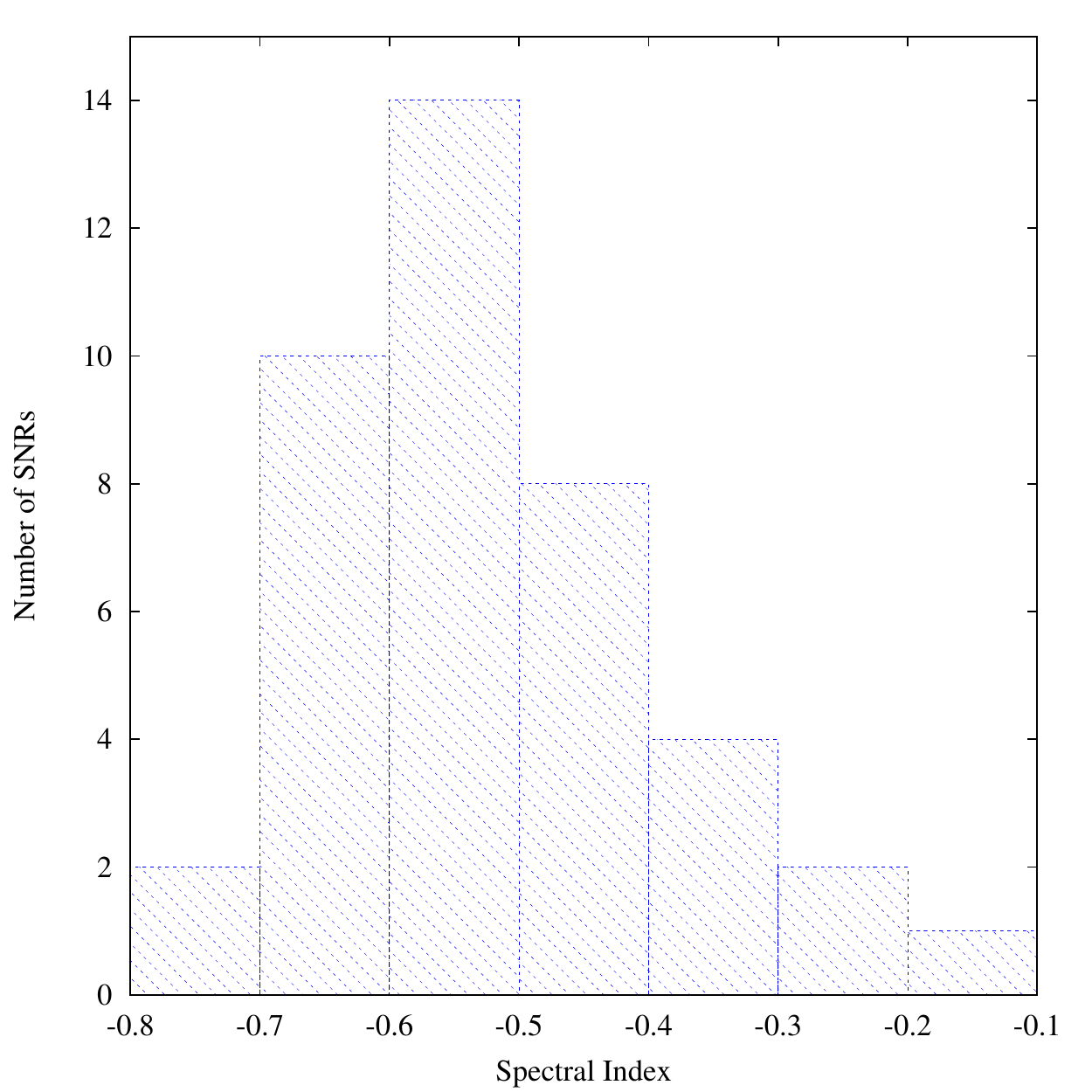}
     \includegraphics[trim=0 0 0 0, width=.485\textwidth]{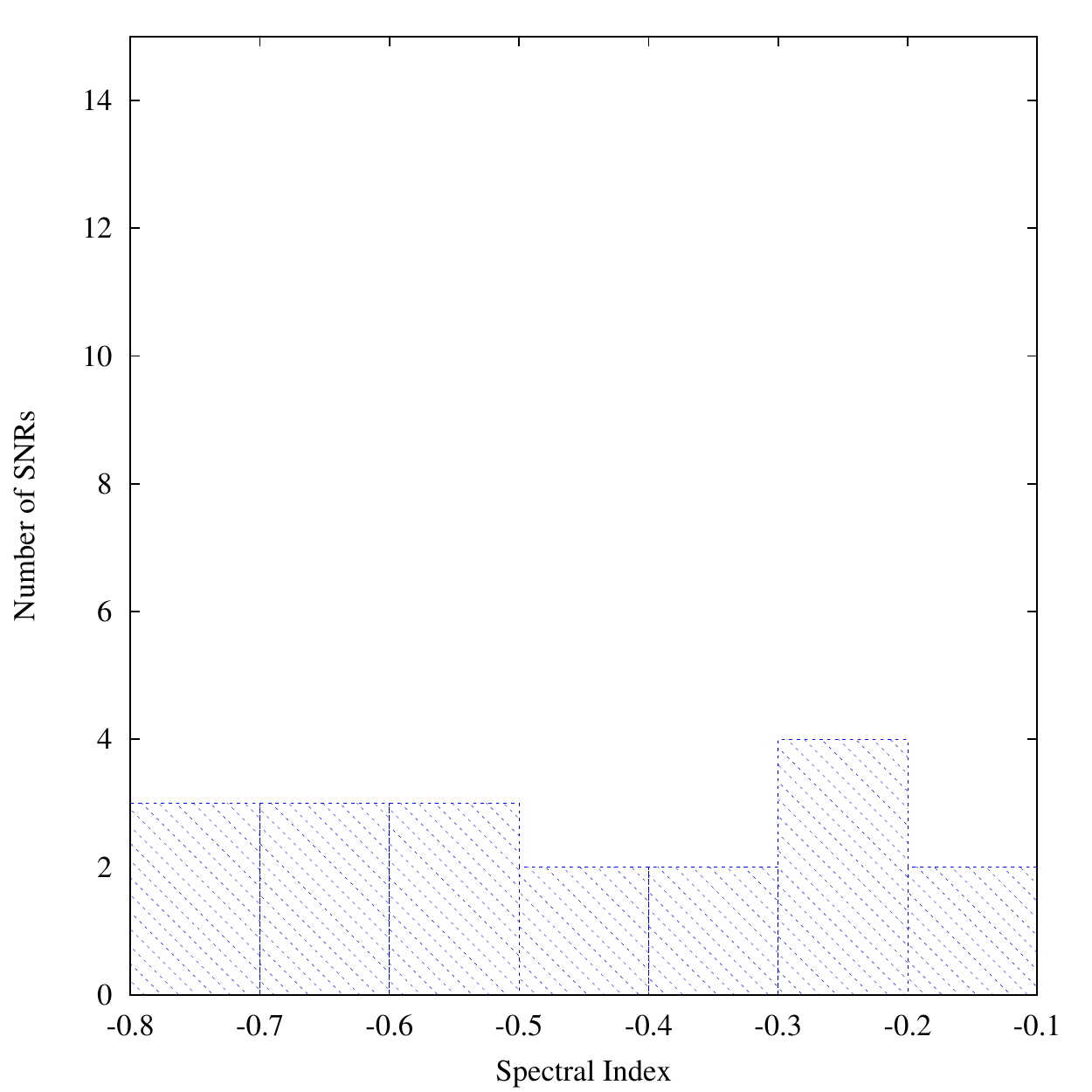}
  \end{center}
  \vspace{-8mm}
  \caption{The radio-continuum spectral index distribution for SNRs in the LMC (left) and SMC (right).}
  \label{fig:si}
\end{figure}

In Fig.~\ref{fig:si} we show the histograms of the radio spectral distribution of SNRs in the LMC and SMC. A mean spectral index of {$-0.52$} is found with a SD of {0.13} for the LMC, while for the SMC we estimate a mean of $-0.47$ with a SD of 0.21. Therefore, both, the LMC and SMC values are inline with the theoretically expected spectral index of $\alpha=-0.5$ (Bell~1978). 

{Filipovi{\'c} et al. (2005)} find a mean radio-continuum spectral index of $-0.63$ (SD=0.43) for confirmed and candidate SNRs in the {SMC}. In our Galaxy, {Clarke \& Caswell (1976)} find a mean of $-0.45$ with (SD$\sim0.15$), while {Xu et al. (2005)} find an average of $-0.5\pm0.25$ using data from the {Green (2004)} catalogue.

\vspace{-3mm}
 \subsection{LMC Remnants from type~Ia SN explosions}

SNRs play a vital role in the physical evolution and chemical enrichment of the ISM. Their precursor, supernovae (SNe), are believed to occur through two main scenarios. The first being core collapse events, which are the explosions of massive stars \mbox{(M $>$ 8--10 $M_{\odot}$)}, and release large quantities of $\alpha$-elements into the ISM. Alternatively, thermonuclear SN (type~Ia) progenitors are less massive and are believed to be the resulting detonation or deflagrations of carbon-oxygen white dwarfs (WDs) that have reached the Chandrasekhar limit ($\sim$1.4~$M_{\odot}$). This may be the rest of a single degenerate system, where a WD in a binary system will accrete matter from a large companion star, or a double degenerate (DD) system, in which two WDs merge (thus exceeding the critical mass) and explode as an SN. The study of type~Ia SNe is not only important due to their use as ``standard candles'' in measuring cosmological distances, but also to ascertain the exact nature of the progenitor system.

\begin{figure}[h!]
  \begin{center}
     \includegraphics[trim=0 0 0 0, width=.16\textwidth]{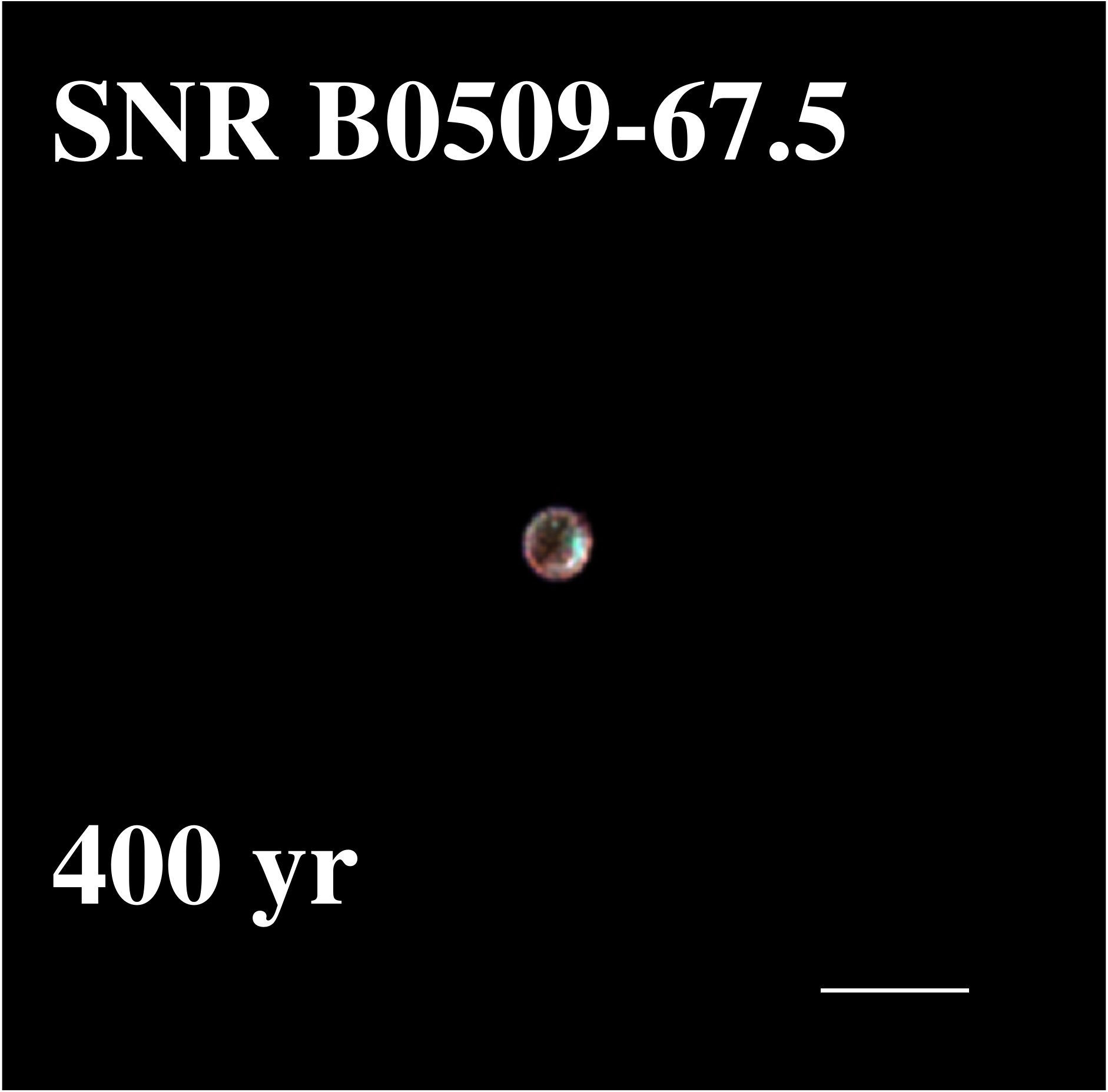}
     \includegraphics[trim=0 0 0 0, width=.16\textwidth]{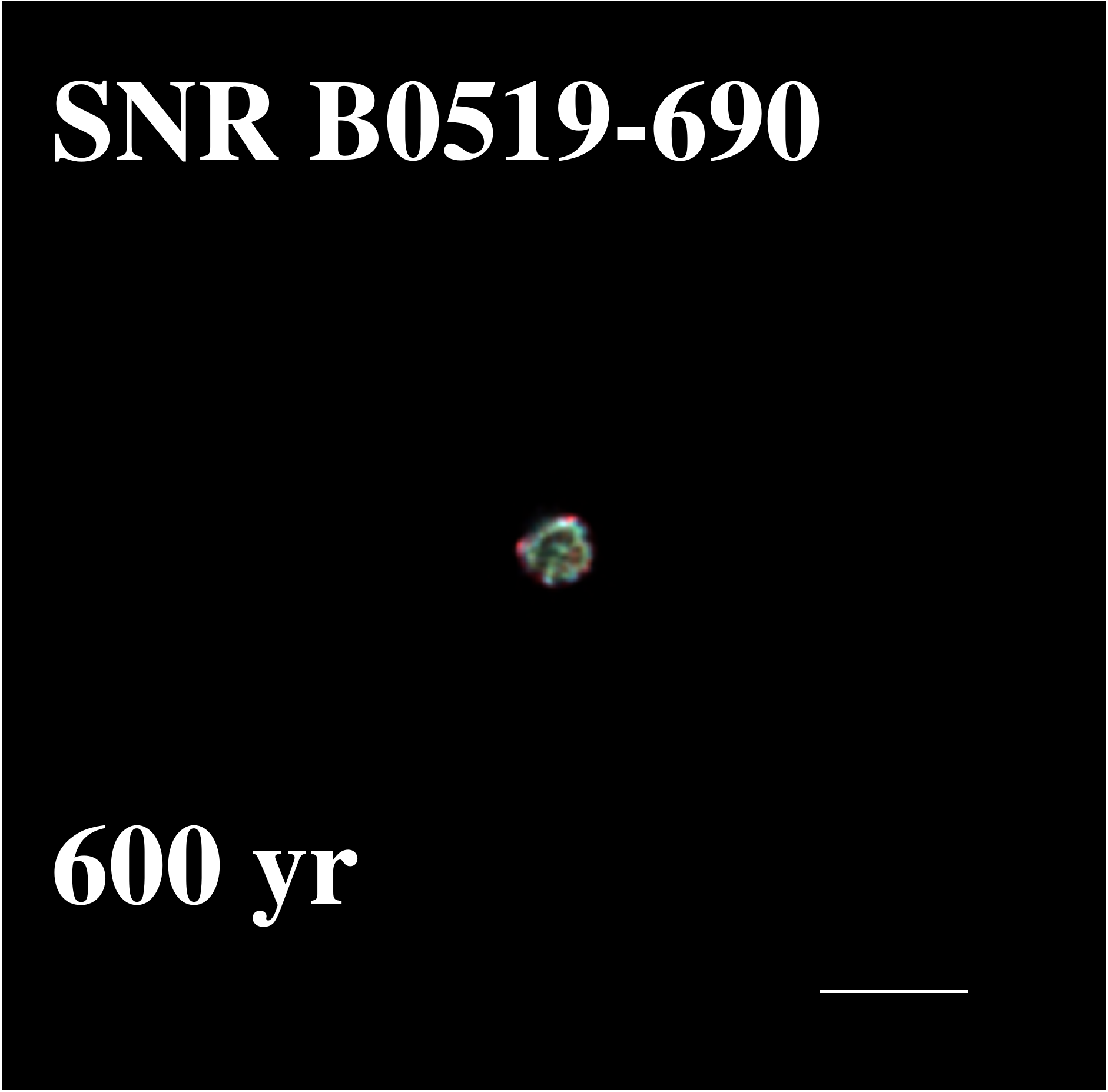}
     \includegraphics[trim=0 0 0 0, width=.16\textwidth]{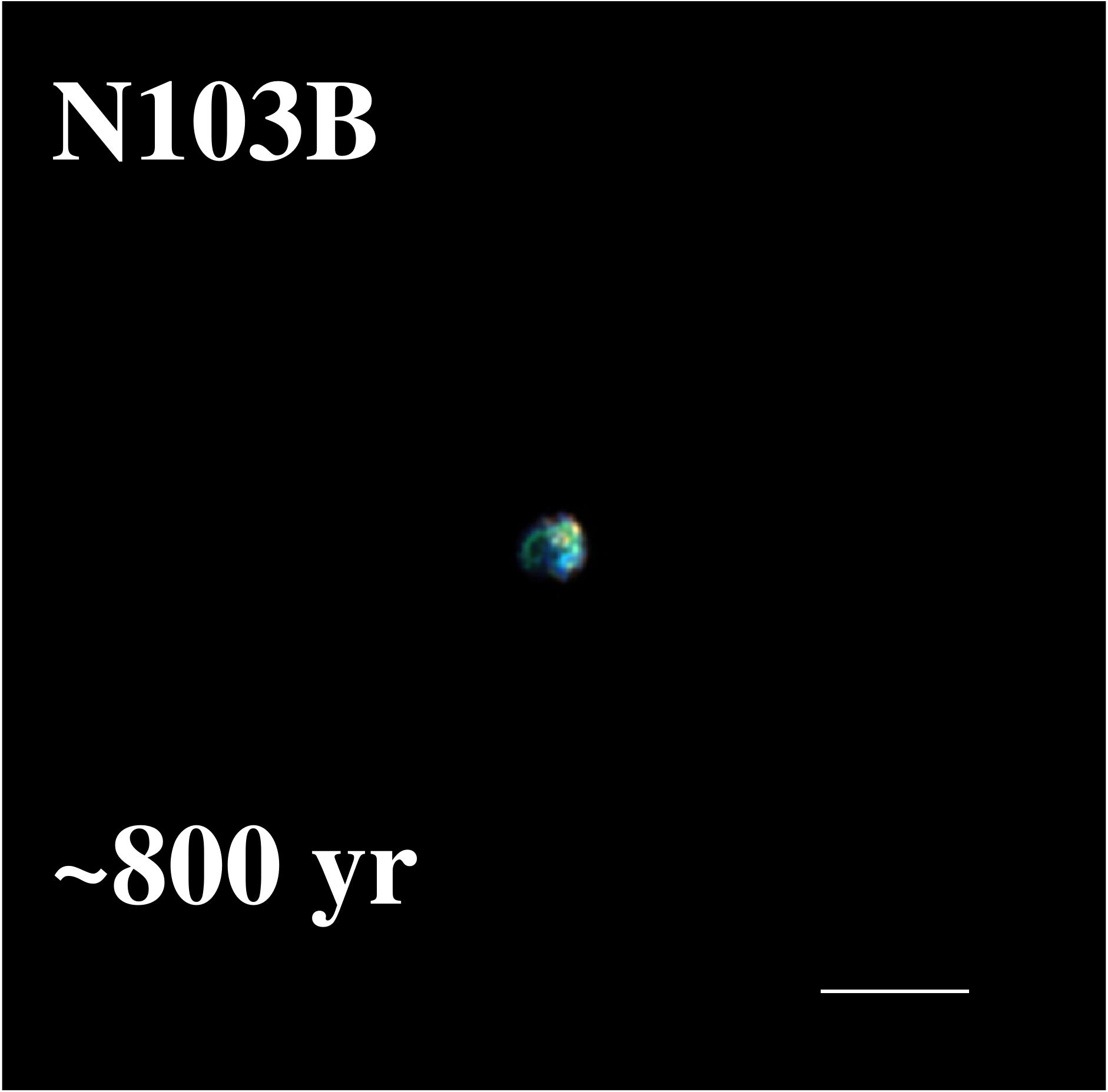}
     \includegraphics[trim=0 0 0 0, width=.16\textwidth]{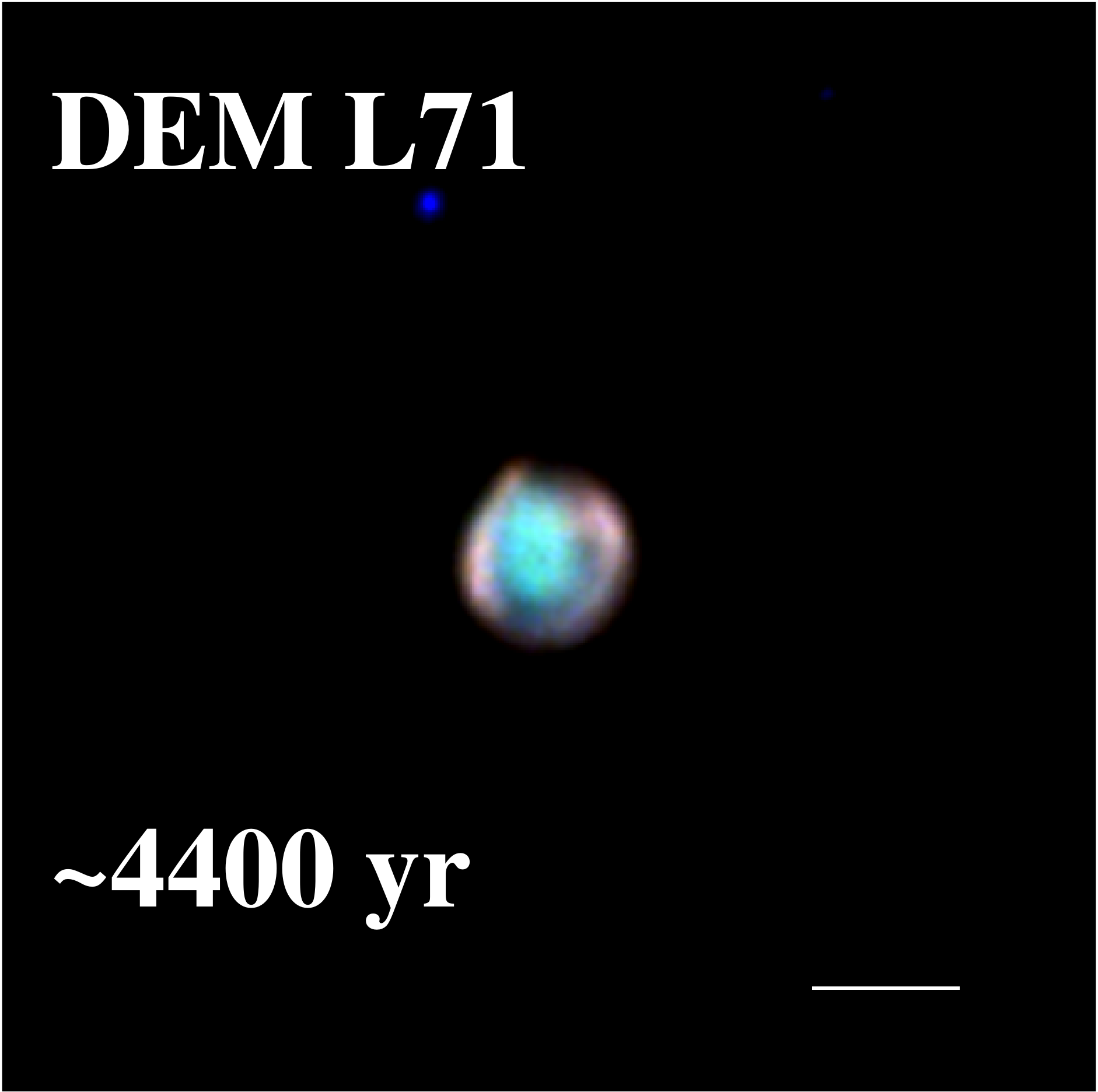}
     \includegraphics[trim=0 0 0 0, width=.16\textwidth]{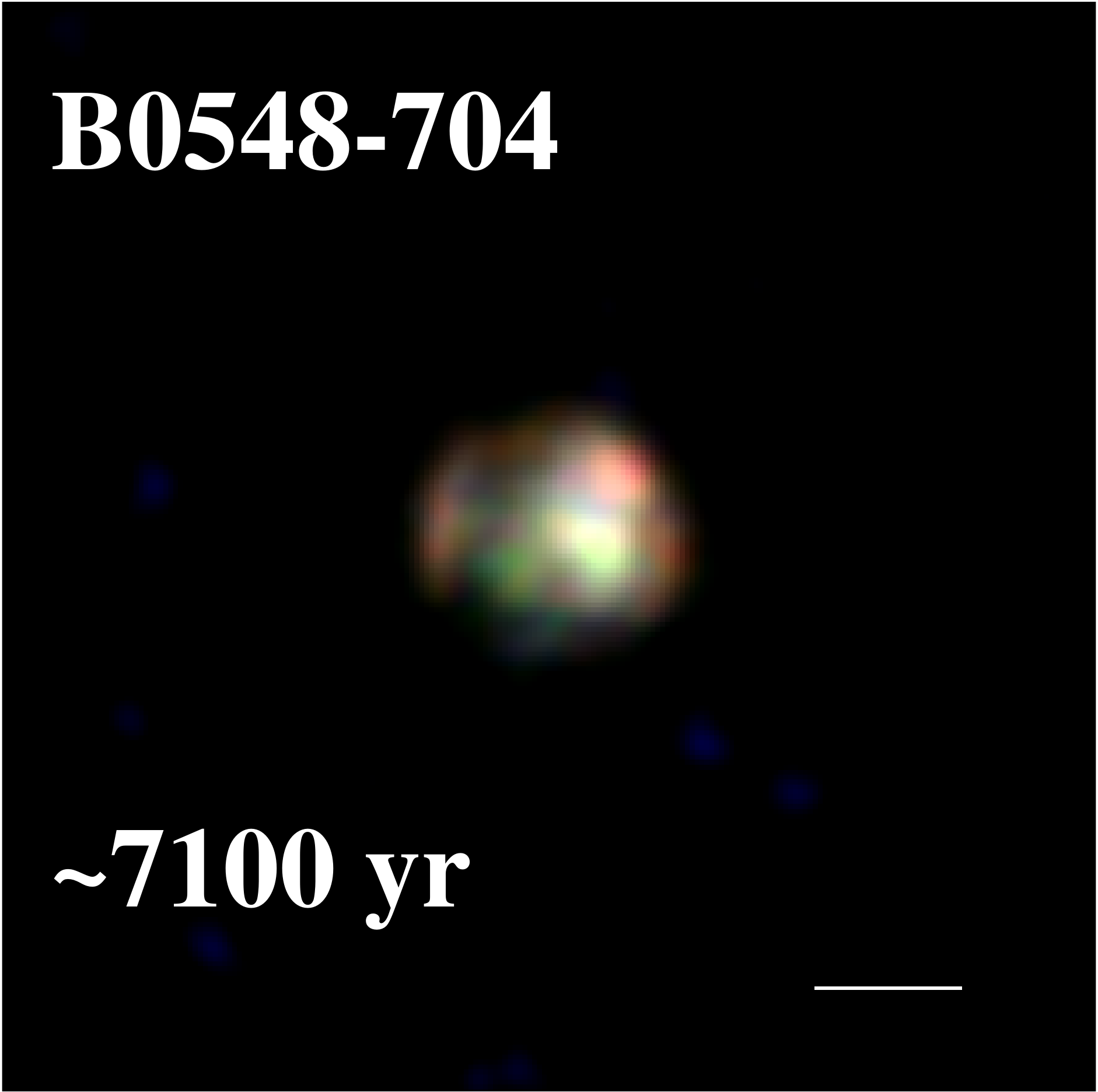}
     \includegraphics[trim=0 0 0 0, width=.16\textwidth]{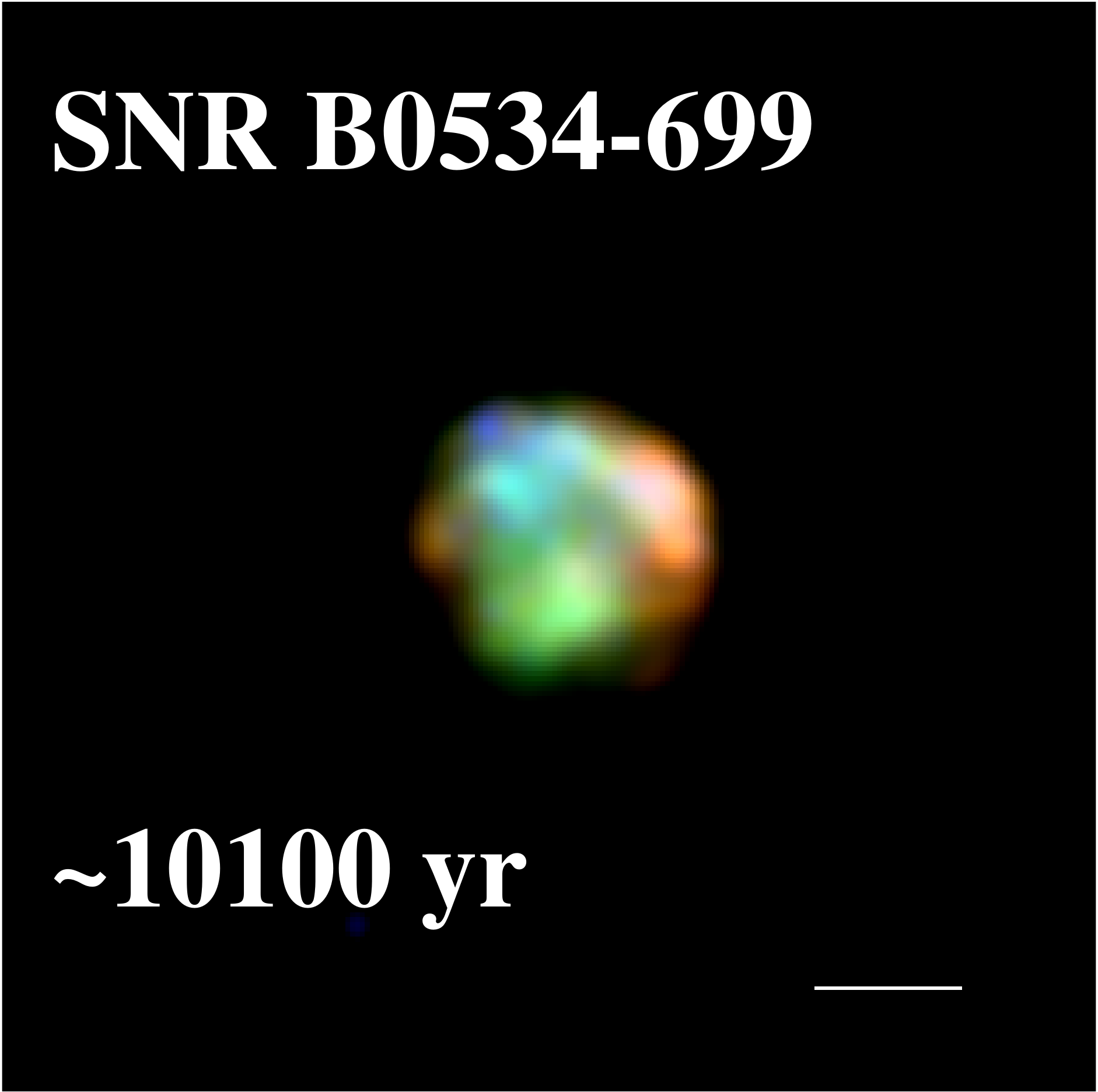}

     \includegraphics[trim=0 0 0 0, width=.16\textwidth]{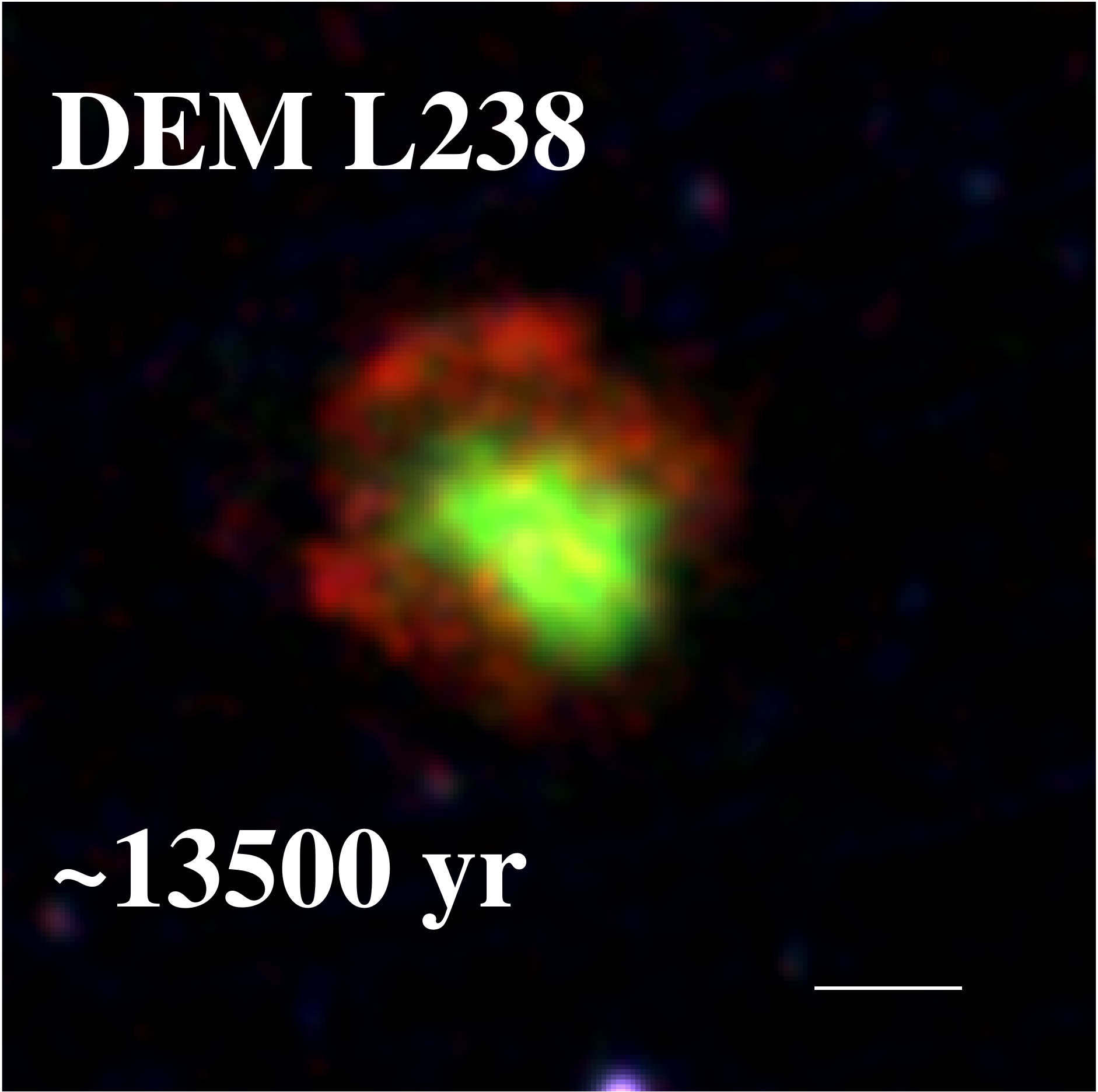}
     \includegraphics[trim=0 0 0 0, width=.16\textwidth]{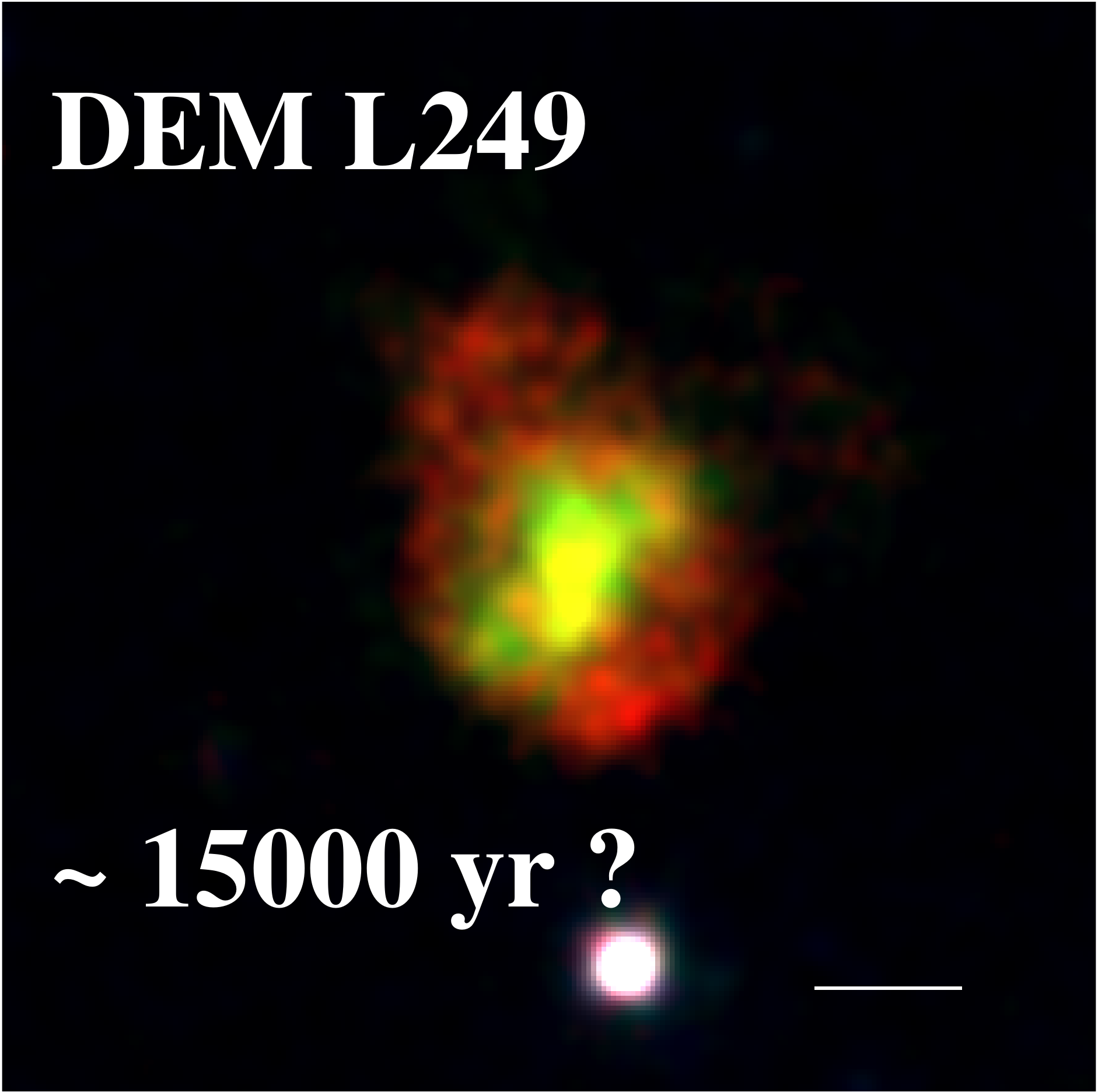}
     \includegraphics[trim=0 0 0 0, width=.16\textwidth]{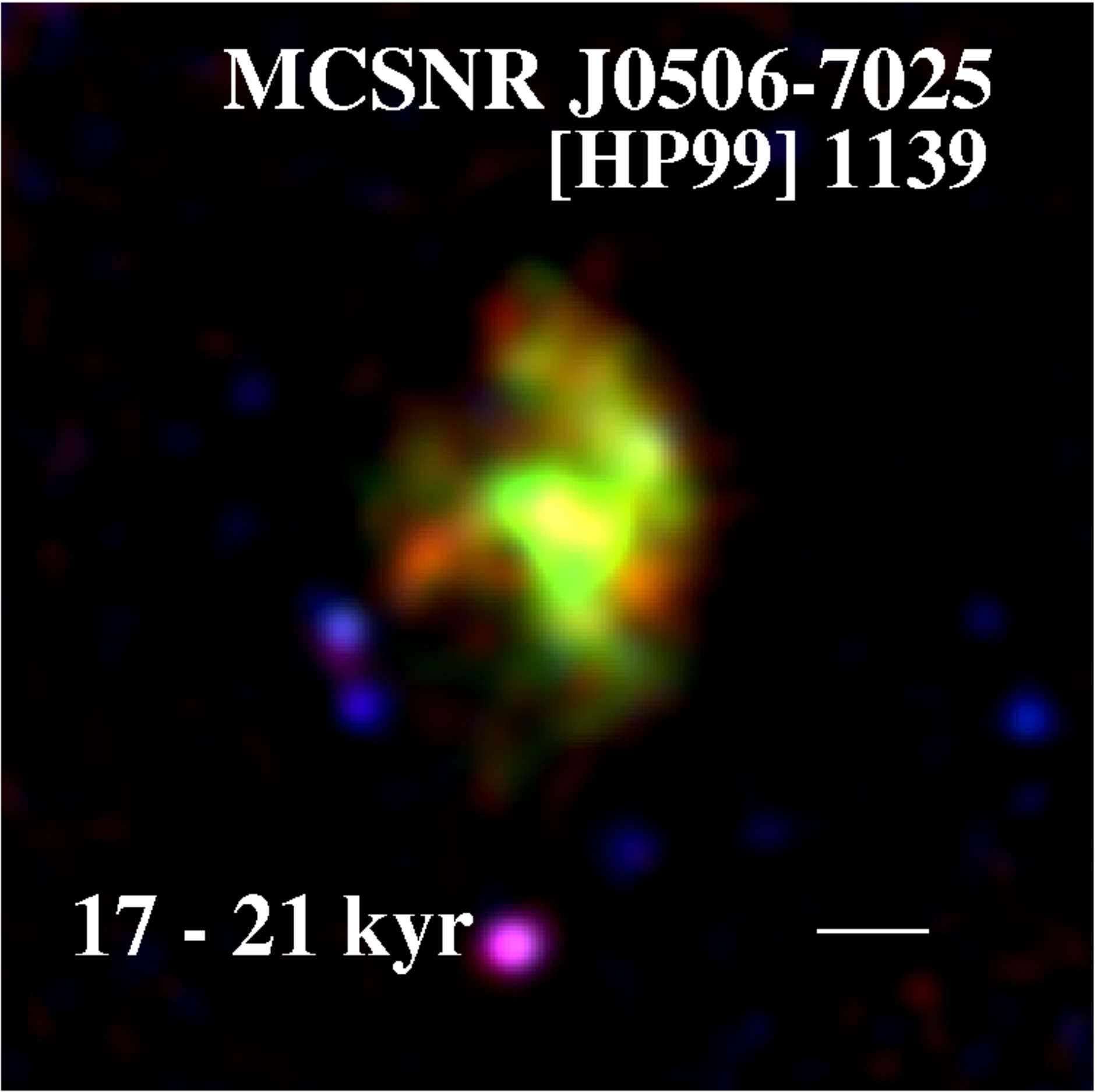}
     \includegraphics[trim=0 0 0 0, width=.16\textwidth]{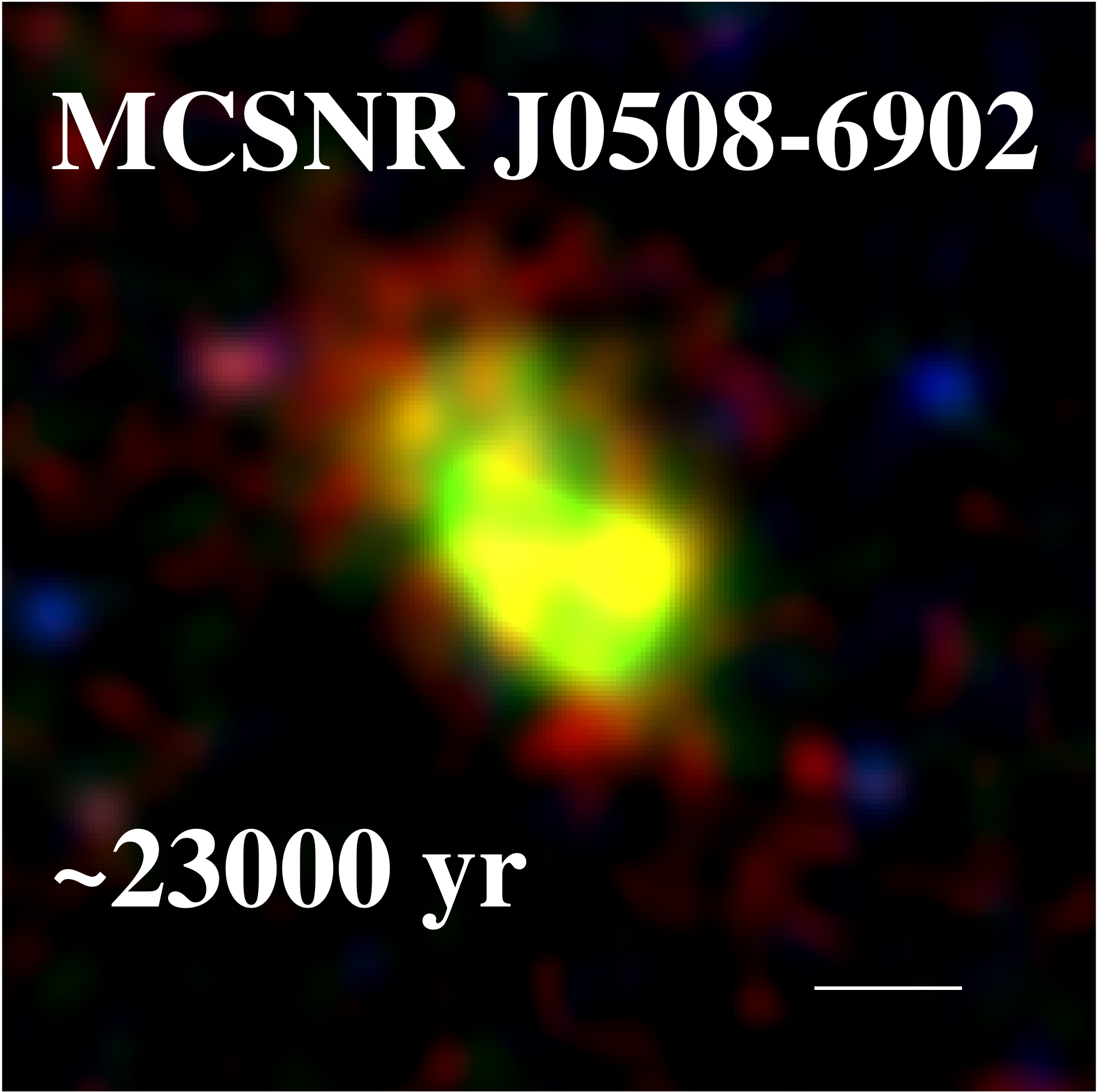}
     \includegraphics[trim=0 0 0 0, width=.16\textwidth]{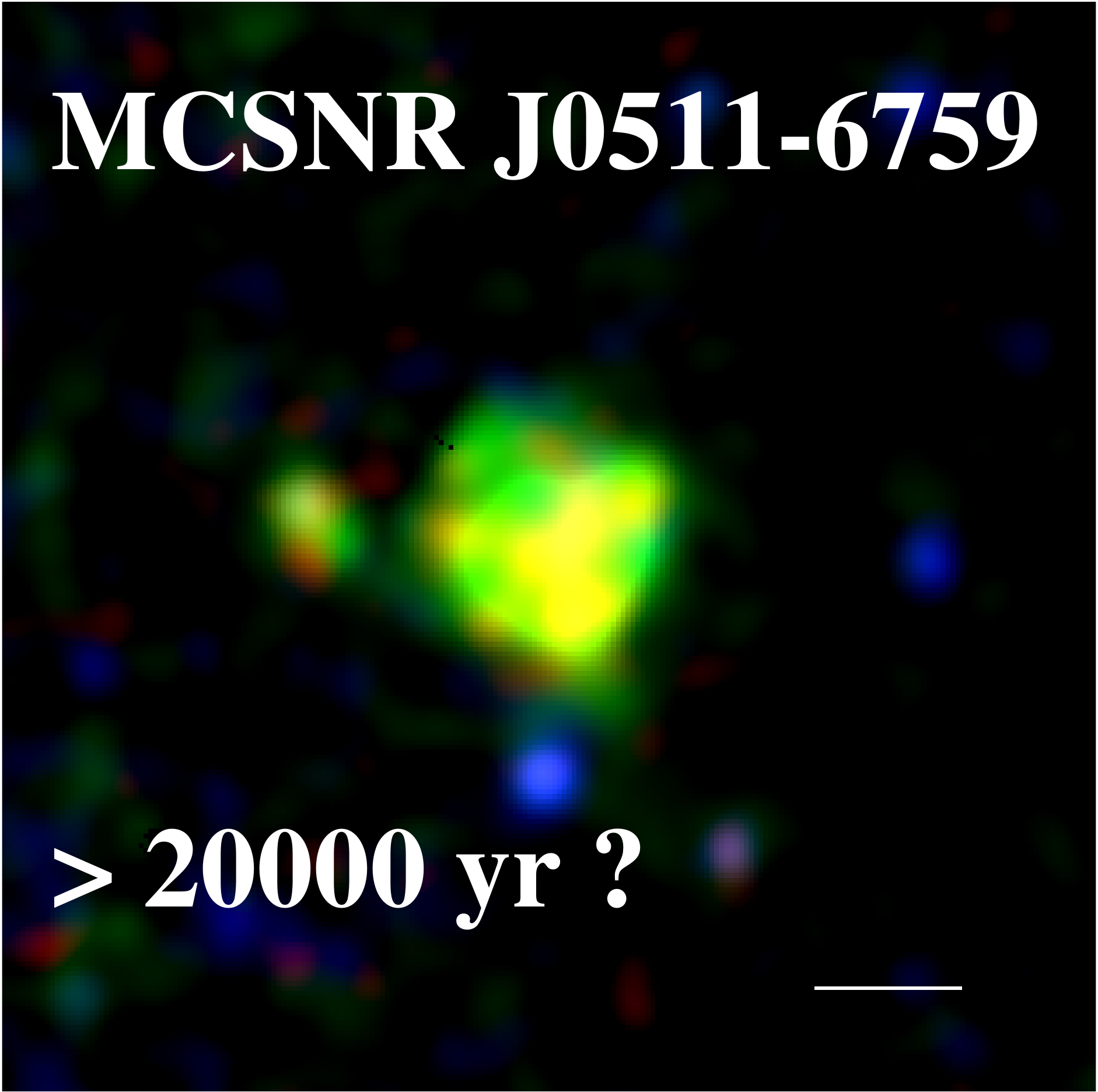}
     \includegraphics[trim=0 0 0 0, width=.16\textwidth]{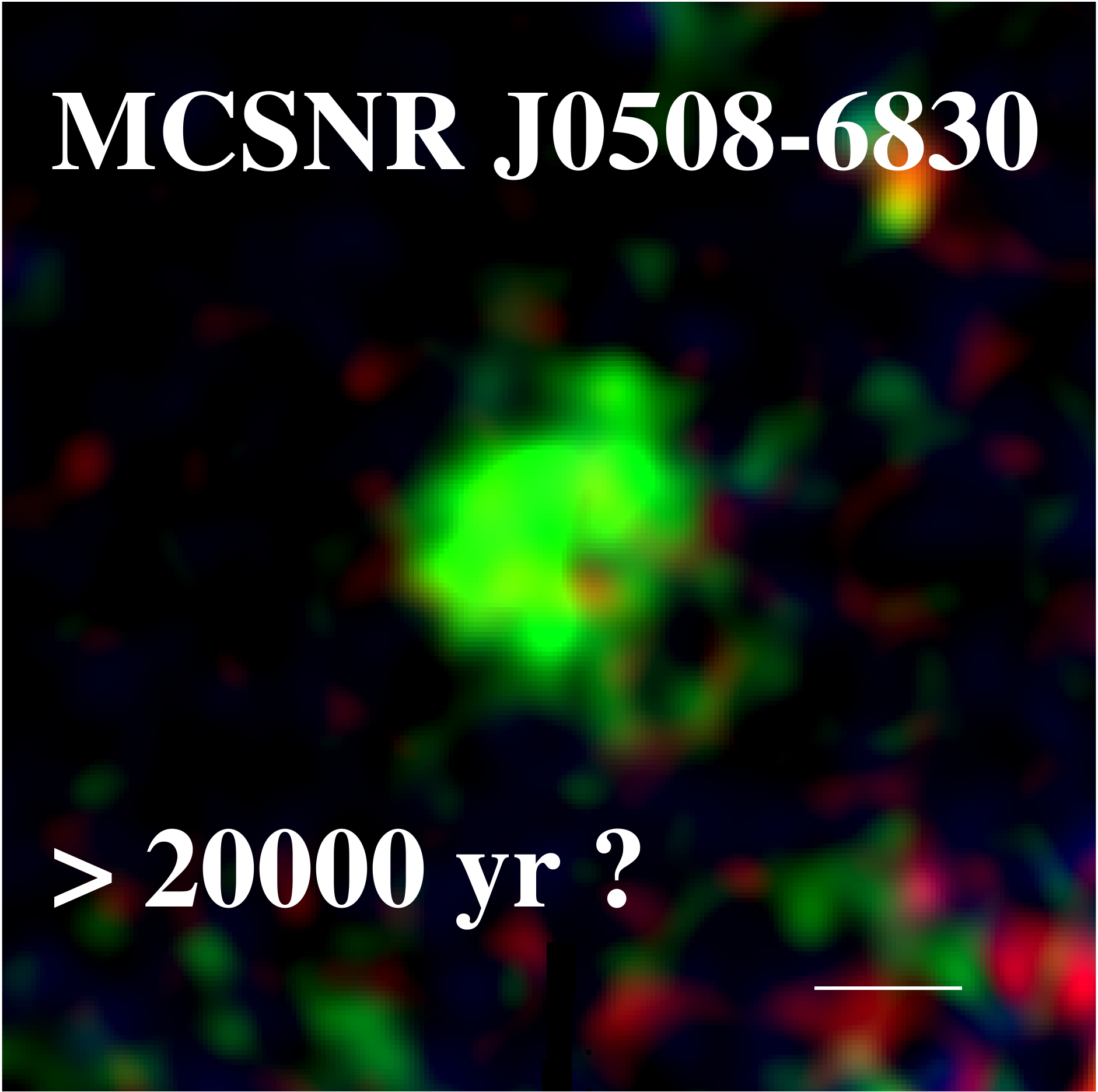}
  \end{center}
  \vspace{-8mm}
  \caption{``Sequence'' of all type~Ia SNRs known in the LMC with available \textit{Chandra} or \textit{XMM-Newton} observations, shown on the same scale (the white bar is 1\arcmin), and sorted by increasing age. We note that with the exception of the younger remnants from who's ages were historic or based on light-echoes, the ages predominately are estimated from X-ray analysis, specifically fitting SNR parameters to the Sedov model, and, therefore, there may be a bias toward remnants following Sedov evolution in this relation. We used \textit{Chandra} data for the three smallest SNRs. In \textit{XMM-Newton} image the red, green and blue components are soft (0.3--0.7~keV), medium (0.7--1.1~keV), and hard (1.1--4.2~keV) X-rays. The medium band is dominated by Fe~L-shell lines, and the iron-rich interior, appearing greenish, is readily distinguished from the softer, fainter shells of the more evolved type~Ia SNRs (second row). Images and caption from Maggi et al. (2016).}
  \label{figtypeIa}
\end{figure}

\begin{figure}[h!]
\vspace{-9mm}
  \begin{center}
    \includegraphics[trim=0 0 0 0, width=.375\textwidth]{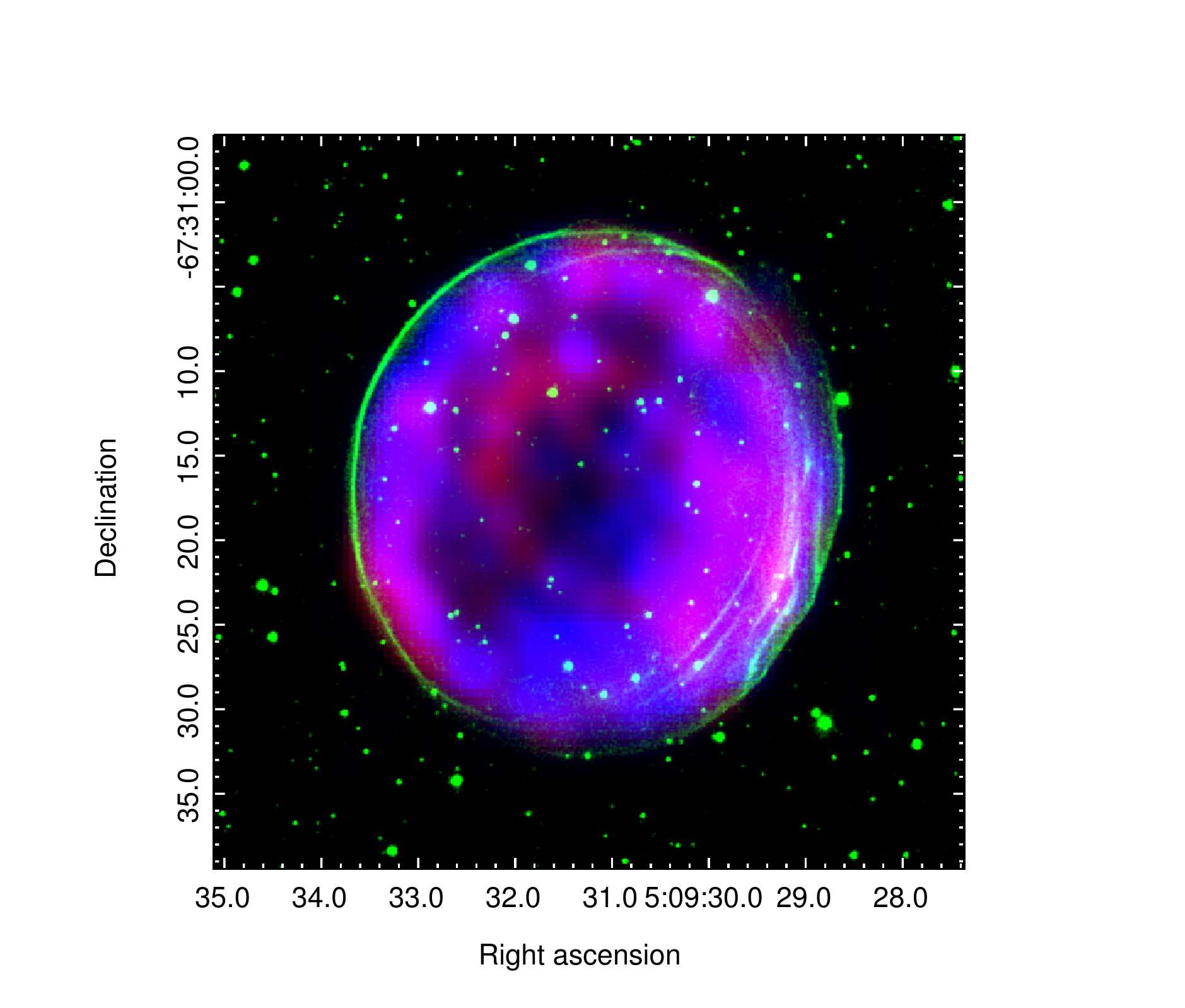}
    \includegraphics[angle=-90,trim=582 140 0 0, width=.23\textwidth]{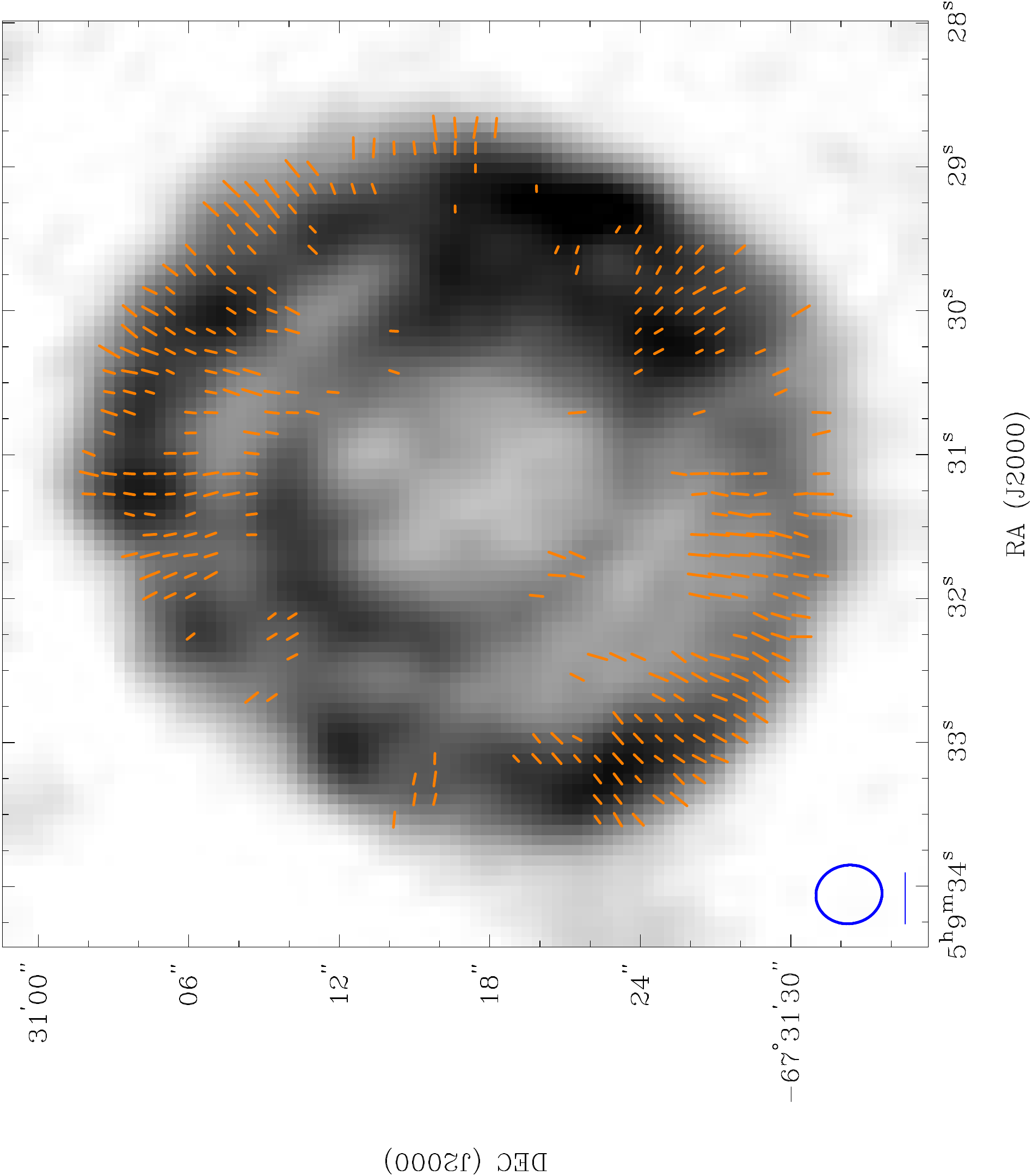}
    \includegraphics[trim=0 0 0 0, width=.375\textwidth]{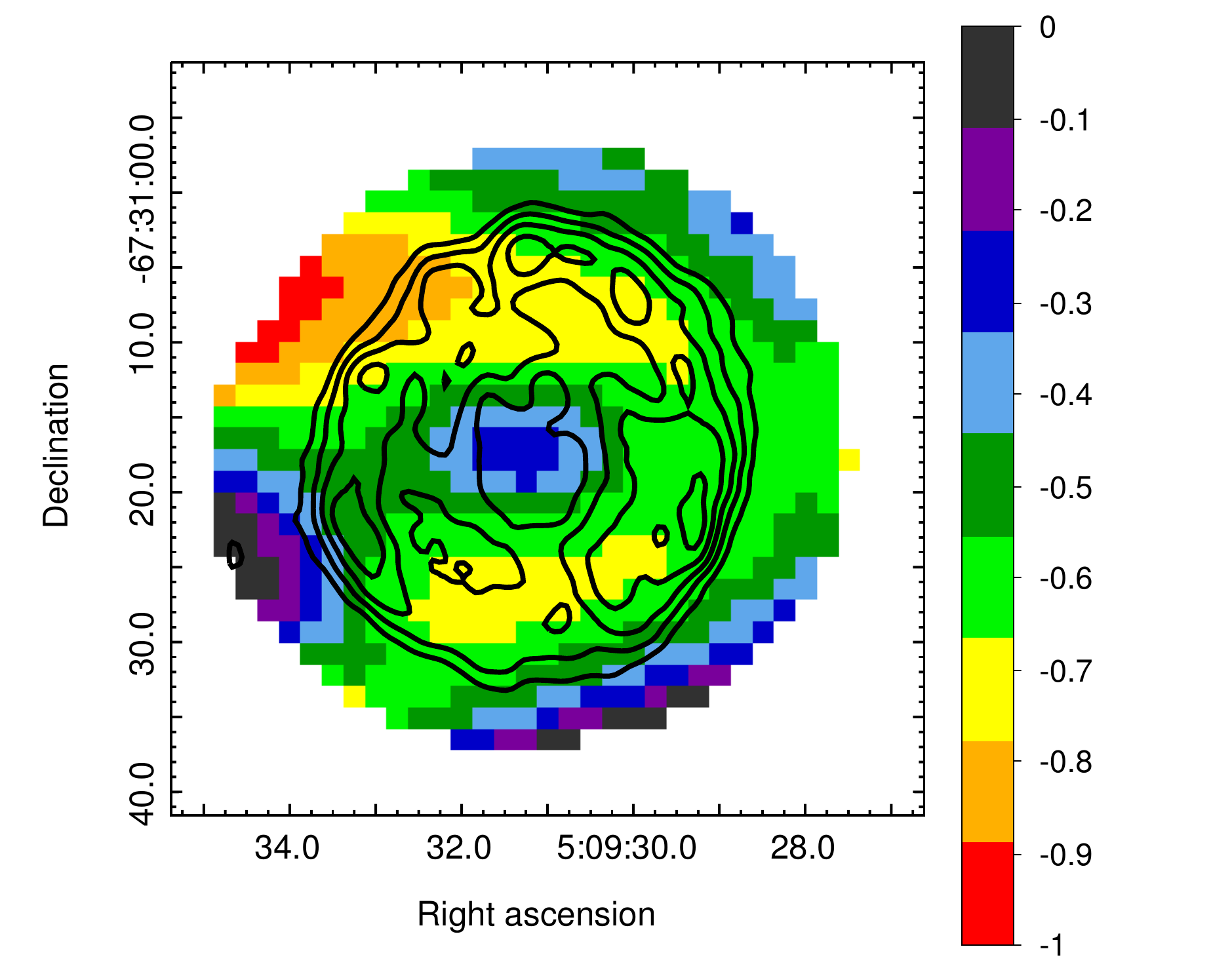}
  \end{center}
  \vspace{-8mm}
  \caption{
  \textit{Left:} Colour composite image of LMC SNR~B0509--67.5 using radio (red), X-ray (blue) and optical (green) observations from the ATCA, {\it Chandra} and Hubble Space Telescope. \textit{Middle:} Intensity image of B0509--67.5 overlaid with polarisation vectors at 6~cm (Bozzetto et al. 2014b; Fig.~6). \textit{Right: }Spectral index map between 13~cm and 6~cm (Bozzetto et al.~2014b; Fig.~5).}
  \label{fig0509}
\end{figure}

Currently (2014), twelve SNRs have been reliably classified as type~Ia in the LMC. Six of these belong to a new sub-class of SNRs containing an iron-rich plasma in their core, including DEM~L238 \& DEM~L249 (Borkowski et al.~2006), MCSNR J0508-6902 (Bozzetto et al.~2014a), MCSNR J0508-6830 and MCSNR J0511-6759 (Maggi et al.~2014) and MCSNR 0506-7025 (Kavanagh et al. 2015c). All twelve of these type~Ia remnants appear prominently in X-ray observations (see Fig.~\ref{figtypeIa}). There have been previous sparse radio observations on a few of these sources, either taken with an insufficient {\it uv} coverage and/or resolution. An example of the benefits from higher resolution observations can be seen in our pilot study on type~Ia LMC~SNR~B0509-67.5 (Bozzetto et al.~2014b), where the inclusion of the long baseline observations made it possible to resolve the radially orientated magnetic field and also, detect a previously unseen central ring within the remnant Fig.~\ref{fig0509}. 

We search for the remnant star (RS) companion (donor star with L$>$10~L$\odot$) in and around all these 12 LMC type~Ia SNRs. Also, Chu et al. (in priv. com.) initial results indicate a lack of expected ``red-giant'' RS within the central region of the SNR. Pagnotta \& Schaefer (2015) also searched two LMC SNRs (SNR~0505--67.9 and SNR~0509--68.7) to identify the centre of the remnant and the 99.73\% containment central region in which any companion star left over from the SN explosion would reside. They concluded that both remnants have a number of potential ex-companion stars near their centres and that either a single or double degenerate scenario may be possible. There are a number of other studies in our Galaxy (for example see Kerzendorf et al. (2014) on Kepler's SNR) that also confirms a lack of RS within type~Ia SNRs. While this may strongly favour a DD scenario, one also should not disregard possible changes in the RS's outer layers which could make them difficult to identify. Most noticeable, Ruiz-Lapuente et al. (2004) found a type G0-G2 star which appears to be the surviving companion of the Tycho SN.

\section{FUTURE WORK}

Our radio-continuum study of SNRs in the MCs are based on ATCA and MOST observations. However, new low-frequency radio-telescopes such as the Murchison Widefield Array (MWA) and Australia Square Kilometre Array Pathfinder (ASKAP) will significantly enhance our knowledge about these objects and will allow us to complete the sample of MCs SNRs. The luminosity-diameter distribution will be used to study the evolution of SNRs in a statistical sense. Also a comparison with the X-ray data will continue the radio -- X-ray comparison. Comprehensive comparison with the optical and IR surveys such as H$\alpha$, [S$\,${\sc ii}], [O$\,${\sc iii}], \textit{Spitzer} and {\it Herschel} surveys are planned as well.

In addition, we will continue our detailed studies on individual SNRs. Some of their characteristics may be related to peculiar properties of the interstellar medium around the explosion site and/or different precursor stars. Theoretical models indicate that changes in density, clumpiness and other properties of the surrounding medium can have a significant effect upon the evolution of an SNR and its emission. However, the collective properties of the SNRs in the MCs studied to date are surprisingly consistent and quite similar to the population of Galactic SNRs despite major differences between these galaxies.

{\bf Acknowledgments}
The ATCA is part of the ATNF, which is funded by the Commonwealth of Australia for operation as a National Facility managed by CSIRO. The scientific results reported in this article are based on observations made by the {\it Chandra} X-ray Observatory and {\it XMM-Newton}. We thank the referee for their excellent comments that improved this manuscript.
%

\references


Bell, A.~R.\ : 1978, 
\journal{Mon. Not. R. Astron. Soc.},  \textbf{182}, 443

Boji{\v c}i{\'c}, I.~S., Filipovi{\'c}, M.~D., Parker, Q.~A., Payne, J.~L., Jones, P.~A., Reid, W., Kawamura, A., Fukui, Y.\ : 2007,  
\journal{Mon. Not. R. Astron. Soc.},  \textbf{378}, 1237 

Borkowski, K.~J., Hendrick, S.~P., Reynolds, S.~P.\ :  
2006, \journal{Astrophys. J.}, \textbf{652}, 1259 

Bozzetto, L.~M., Filipovi{\'c}, M.~D., Crawford, E.~J., Boji{\v c}i\'c, I.~S., Payne, J.~L., Medik, A., Wardlaw, B., De Horta, A.~Y.\ : 2010,  
\journal{Serb. Astron. J.}, \textbf{181}, 43 

Bozzetto, L.~M., Filipovi{\'c}, M.~D., Crawford, E.~J., Haberl, F., Sasaki, M., Uro{\v s}evi{\'c}, D., Pietsch, W., Payne, J.~L., De Horta, A.~Y., Stupar, M., Tothill, N.~F.~H., Dickel, J., Chu, Y.-H., Gruendl, R.\ : 2012a,  
\journal{Mon. Not. R. Astron. Soc.}, \textbf{420}, 2588 

Bozzetto, L.~M., Filipovi{\'c}, M.~D., Crawford, E.~J., Payne, J.~L., De Horta, A.~Y., Stupar, M.\ : 2012b,  
\journal {Rev. Mex. Astron. Astrofis.}, \textbf{48}, 41 

Bozzetto, L.~M., Filipovi{\'c}, M.~D., Uro{\v s}evi\'c, D., Crawford, E.~J.\ : 2012c,  
\journal{Serb. Astron. J.},  \textbf{185}, 25 

Bozzetto, L.~M., Filipovi{\'c}, M.~D., Crawford, E.~J., Sasaki, M., Maggi, P., Haberl, F., Uro{\v s}evi{\'c}, D., Payne, J.~L., De Horta, A.~Y., Stupar, M., Gruendl, R., Dickel, J.\ : 2013,  
\journal{Mon. Not. R. Astron. Soc.}, \textbf{432}, 2177 

Bozzetto, L.~M., Kavanagh, P.~J., Maggi, P., Filipovi{\'c}, M.~D., Stupar, M., Parker, Q.~A., Reid, W.~A., Sasaki, M., Haberl, F., Uro{\v s}evi{\'c}, D., Dickel, J., Sturm, R., Williams, R., Ehle, M., Gruendl, R., Chu, Y.-H., Points, S., Crawford, E.~J.\ : 2014a,  
\journal{Mon. Not. R. Astron. Soc.},  \textbf{439}, 1110 

Bozzetto, L.~M., Filipovi{\'c}, M.~D., Uro{\v s}evi{\'c}, D., Kothes, R., Crawford, E.~J.\ : 2014b,  
\journal{Mon. Not. R. Astron. Soc.}, \textbf{440}, 3220 

Bozzetto, L.~M., Filipovi{\'c}, M.~D.\ : 2014c,  
\journal{Astrophys. Space Sci.},  \textbf{351}, 207 

Bozzetto, L.~M., Filipovi{\'c}, M.~D.\ : 2015,  
\journal{Pub. Korean Astro. Soc.}, \textbf{30}, 149 

Brantseg, T., McEntaffer, R.~L., Bozzetto, L.~M., Filipovi{\'c}, M., Grieves,  N.\ : 2014,  
\journal{Astrophys. J.}, \textbf{780}, 50 

{\v C}ajko, K.~O., Crawford, E.~J., Filipovi{\'c}, M.~D.\ : 2009,  
\journal{Serb. Astron. J.},  \textbf{179}, 55 

Clark, D.~H., Caswell, J.~L.\ : 1976,  
\journal{Mon. Not. R. Astron. Soc.}, \textbf{174}, 267 

Crawford, E.~J., Filipovi{\'c}, M.~D., Haberl, F., Pietsch, W., Payne, J.~L., De Horta, A.~Y.\ : 2010,  
\journal{Astron. Astrophys.}, \textbf{518}, A35 

Crawford, E.~J., Filipovi\'c, M.~D., De Horta, A.~Y., Wong, G.~F., Tothill, N.~F.~H., Dra{\v s}kovi\'c, D., Collier, J.~D., Galvin, T.~J.\ : 2011,  
\journal{Serb. Astron. J.},  \textbf{183}, 95 

Crawford, E.~J., Filipovi{\'c}, M.~D., McEntaffer, R.~L., Brantseg, T., Heitritter, K., Roper, Q., Haberl, F., Uro{\v s}evi{\'c}, D.\ : 2014,  
\journal{Astron. J.},  \textbf{148}, 99 

De Horta, A.~Y., Filipovi{\'c}, M.~D., Bozzetto, L.~M., Maggi, P., Haberl, F., Crawford, E.~J., Sasaki, M., Uro{\v s}evi{\'c}, D., Pietsch, W., Gruendl, R., Dickel, J., Tothill, N.~F.~H., Chu, Y.-H., Payne, J.~L., Collier, J.~D.\ : 2012,  
\journal{Astron. Astrophys.},  \textbf{540}, A25 

De Horta, A.~Y., Sommer, E.~R., Filipovi{\'c}, M.~D., O'Brien, A., Bozzetto, L.~M., Collier, J.~D., Wong, G.~F., Crawford, E.~J., Tothill, N.~F.~H., Maggi, P., Haberl, F.\ : 2014,  
\journal{Astron. J.}, \textbf{147}, 162 

di Benedetto, G.~P.\ : 2008,
\journal{Mon. Not. R. Astron. Soc.}, \textbf{390}, 1762 

Dickel, J.~R., McIntyre, V.~J., Gruendl, R.~A., Milne, D.~K.\ : 2005,  
\journal{Astron. J.},  \textbf{129}, 790 

Dopita, M.~A., Blair, W.~P., Long, K.~S., Mutchler, M., Whitmore, B.~C., Kuntz, K.~D., Balick, B., Bond, H.~E., Calzetti, D., Carollo, M., Disney, M., Frogel, J.~A., O'Connell, R., Hall, D., Holtzman, J.~A., Kimble, R.~A., MacKenty, J., McCarthy, P., Paresce, F., Saha, A., Silk, J., Sirianni, M., Trauger, J., Walker, A.~R., Windhorst, R., Young, E.\ : 2010,  
\journal{Astrophys. J.}, \textbf{710}, 964 

Filipovi\'c, M.~D., Haynes, R.~F., White, G.~L., Jones, P.~A., Klein, U., Wielebinski, R.\ : 1995,  
\journal{Astron. Astrophys. Suppl. Series}, \textbf{111}, 311 

Filipovi\'c, M.~D., Jones, P.~A., White, G.~L., Haynes, R.~F., Klein, U.,  Wielebinski, R.\ : 1997,  
\journal{Astron. Astrophys. Suppl. Series}, \textbf{121}, 321 

Filipovi\'c, M.~D., Haynes, R.~F., White, G.~L., Jones, P.~A.\ : 1998,  
\journal{Astron. Astrophys. Suppl. Series},  \textbf{130}, 421 

Filipovi{\'c}, M.~D., Bohlsen, T., Reid, W., Staveley-Smith, L., Jones, P.~A., Nohejl, K., Goldstein, G.\ : 2002,  
\journal{Mon. Not. R. Astron. Soc.},  \textbf{335}, 1085 

Filipovi{\'c}, M.~D., Payne, J.~L., Reid, W., Danforth, C.~W., Staveley-Smith, L., Jones, P.~A., White, G.~L.\ : 2005,  
\journal{Mon. Not. R. Astron. Soc.},  \textbf{364}, 217 

Filipovi{\'c}, M.~D., Haberl, F., Winkler, P.~F., Pietsch, W., Payne, J.~L., Crawford, E.~J., De Horta, A.~Y., Stootman, F.~H., Reaser, B.~E.\ :  2008,  
\journal{Astron. Astrophys.},  \textbf{485}, 63 

Green, D.~A.\ : 2004,  
\journal{Bull. Astr. Soc. India},  \textbf{32}, 335 

Grondin, M.-H., Sasaki, M., Haberl, F., Pietsch, W., Crawford, E.~J., Filipovi{\'c}, M.~D., Bozzetto, L.~M., Points, S., Smith, R.~C.\ : 2012,  
\journal{Astron. Astrophys.}, \textbf{539}, A15 

Haberl, F., Sturm, R., Ballet, J., Bomans, D.~J., Buckley, D.~A.~H., Coe, M.~J., Corbet, R., Ehle, M., Filipovi\'c, M.~D., Gilfanov, M., Hatzidimitriou, D., La Palombara, N., Mereghetti, S., Pietsch, W., Snowden, S., Tiengo, A.\ : 2012,  
\journal{Astron. Astrophys.},  \textbf{545}, A128 

Hughes, A., Staveley-Smith, L., Kim, S., Wolleben, M., Filipovi{\'c}, M.\ :  2007,  
\journal{Mon. Not. R. Astron. Soc.},  \textbf{382},  543 

Kavanagh, P.~J., Sasaki, M., Points, S.~D., Filipovi{\'c}, M.~D., Maggi, P., Bozzetto, L.~M., Crawford, E.~J., Haberl, F., Pietsch, W.\ : 2013,  
\journal{Astron. Astrophys.},  \textbf{549}, A99 

Kavanagh, P.~J., Sasaki, M., Bozzetto, L.~M., Filipovi{\'c}, M.~D., Points, S.~D., Maggi, P., Haberl, F.\ : 2015a,  
\journal{Astron. Astrophys.},  \textbf{573}, A73 

Kavanagh, P.~J., Sasaki, M., Whelan, E. T., Maggi, P., Haberl, F., Bozzetto, L.~M., Filipovi{\'c}, M.~D., Crawford, E.~J. \ : 2015b,  
\journal{Astron. Astrophys.},  \textbf{579}, A63 

Kavanagh, P.~J., Sasaki, M., Bozzetto, L.~M., Points, S.~D., Filipovi{\'c}, M.~D., Maggi, P., Haberl, F., Crawford, E.~J. \ : 2015c,  
\journal{Astron. Astrophys.},  \textbf{583}, A121 

Kavanagh, P.~J., Sasaki, M., Bozzetto, L.~M., Points, S.~D., Crawford, E.~J., Dickel, J., Filipovi{\'c}, M.~D., Haberl, F., Maggi, P., Whelan, E. T. \ : 2016,  
\journal{Astron. Astrophys.},  \textbf{586}, A4 

Kerzendorf, W. E., Childress, M., Scharwachter, J., Do, T., Schmidt, B. P.\ : 2014,
\journal{Astrophys. J.},  \textbf{782}, 27 

Laki\'cevi\'c, M., van Loon, J. T., Meixner, M., Gordon, K., Bot, C., Roman-Duval, J., Babler, B., Bolatto, A., Engelbracht, C., Filipovi{\'c}, M.~D., Hony, S., Indebetouw, R., Misselt, K., Montiel, E., Okumura, K., Panuzzo, P., Patat, F., Sauvage, M., Seale, J., Sonneborn, G., Temim, T., Uro{\v s}evi{\'c}, D., Zanardo, G.\ : 2015, 
\journal{Astrophys. J.}, \textbf{799}, 50L

Lee, J.~H., Lee, M.~G.\ : 2014,  
\journal{Astrophys. J.},  \textbf{786},  130 

Long, K.~S., Blair, W.~P., Winkler, P.~F., Becker, R.~H., Gaetz, T.~J., Ghavamian, P., Helfand, D.~J., Hughes, J.~P., Kirshner, R.~P., Kuntz, K.~D., McNeil, E.~K., Pannuti, T.~G., Plucinsky, P.~P., Saul, D., T{\"u}llmann, R., Williams, B.\ : 2010,  
\journal{Astrophys. J. Suppl. Series},  \textbf{187}, 495 

Maggi, P., Haberl, F., Kavanagh, P.~J., Points, S.~D., Dickel, J.,  Bozzetto, L.~M., Sasaki, M., Chu, Y.-H., Gruendl, R.~A., Filipovi{\'c}, M.~D., Pietsch, W.\ : 2014,  
\journal{Astron. Astrophys.},  \textbf{561},  A76 

Maggi, P., Haberl, F., Kavanagh, P.~J., Sasaki, M., Bozzetto, L.~M., Filipovi{\'c}, M.~D., Vasilopoulos, G., Pietsch, W., Points, S. D., Chu, Y.-H., Dickel, J., Ehle, M., Williams, R., Greiner, J.\ : 2016,  
\journal{Astron. Astrophys.},  \textbf{585},  A162

Maitra, C., Ballet, J., Filipovi{\'c}, M.~D., Haberl, F., Tiengo, A., Grieve, K., Roper, Q. \ : 2015,
\journal{Astron. Astrophys.},  \textbf{584}, A41 

Mills, B.~Y., Turtle, A.~J.\ : 1984,  
IN: Structure and evolution of the Magellanic Clouds; Proceedings of the Symposium, Tuebingen, West Germany, September 5-8, 1983 (A85-26576 11-90). Dordrecht, D. Reidel Publishing Co.,  \textbf{Vol.~108}, 283 

Owen, R.~A., Filipovi{\'c}, M.~D., Ballet, J., Haberl, F., Crawford, E.~J., Payne, J.~L., Sturm, R., Pietsch, W., Mereghetti, S., Ehle, M., Tiengo, A., Coe, M.~J., Hatzidimitriou, D., Buckley, D.~A.~H.\ : 2011,  
\journal{Astron. Astrophys.},  \textbf{530}, A132 

Pagnotta A., Schaefer B.~E.\ 2015, 
\journal{Astrophys. J.},  \textbf{799}, 101

Payne, J.~L., Filipovi{\'c}, M.~D., Reid, W., Jones, P.~A., Staveley-Smith, L., White, G.~L.\ : 2004,  
\journal{Mon. Not. R. Astron. Soc.},  \textbf{355}, 44 

Payne, J.~L., White, G.~L., Filipovi{\'c}, M.~D., Pannuti, T.~G.\ : 2007,  
\journal{Mon. Not. R. Astron. Soc.},  \textbf{376}, 1793 

Payne, J.~L., White, G.~L., Filipovi{\'c}, M.~D.\ : 2008,  
\journal{Mon. Not. R. Astron. Soc.},  \textbf{383}, 1175 

Reid, W. A., Stupar, M., Bozzetto, L. M., Parker, Q. A., Filipovi{\'c}, M.~D.\ : 2015,  
\journal{Mon. Not. R. Astron. Soc.},  \textbf{454}, 991 

Roper, Q., McEntaffer, R. L., DeRoo, C., Filipovi{\'c}, M.~D., Wong, G. F., Crawford, E. J.\ : 2015,
\journal{Astrophys. J.},  \textbf{803}, 106

Ruiz-Lapuente, P., Comeron, F., Mendez, J., Canal, R., Smartt, S. J., Filippenko, A. V., Kurucz, R. L., Chornock, R., Foley, R. J., Stanishev, V., Ibata, R.\ : 2004,  
\journal{Nature}, \textbf{420}, 1069

Smith C., Points S., Winkler P.~F.\ : 2006,  
\journal{NOAO Newsletter},  \textbf{85}, 6

Warth, G., Sasaki, M., Kavanagh, P., Filipovi{\'c}, M.~D., Points, S. D.; Bozzetto, L.~M.\ : 2014,
\journal{Astron. Astrophys.},  \textbf{567}, A136

Wong, G.~F., Filipovi{\'c}, M.~D., Crawford, E.~J., De Horta, A.~Y., Galvin, T., Dra{\v s}kovi{\'c}, D., and Payne, J.~L.\ : 2011a, 
\journal{Serb. Astron. J.},  \textbf{182}, 43 

Wong, G.~F., Filipovi{\'c}, M.~D., Crawford, E.~J., Tothill, N.~F.~H., De Horta, A.~Y., Dra{\v s}kovi{\'c}, D., Galvin, T.~J., Collier, J.~D., and Payne, J.~L.\ : 2011b, 
\journal{Serb. Astron. J.},  \textbf{183}, 103 

Wong, G.~F., Crawford, E.~J., Filipovi{\'c}, M.~D., De Horta, A.~Y., Tothill, N.~F.~H., Collier, J.~D., Dra{\v s}kovi{\'c}, D., Galvin, T.~J., Payne, J.~L.\  : 2012, 
\journal{Serb. Astron. J.},  \textbf{184}, 93 

Xu, J.-W., Zhang, X.-Z., Han, J.-L.\ : 2005,  
\journal{Chin. J. Astron. Astrophys.}  \textbf{5}, 165 

\endreferences

\end{document}